\documentclass[submission,copyright,creativecommons]{eptcs}
\providecommand{\event}{ThEdu 24} 

\usepackage{iftex}

\ifpdf
  \usepackage{underscore}         
  \usepackage[T1]{fontenc}        
\else
  \usepackage{breakurl}           
\fi

\usepackage{hyperref}
\usepackage{amssymb, amsmath, amsthm, mathtools}
\usepackage{amsfonts, bbold, dsfont}

\DeclarePairedDelimiterXPP\norm[1]{}\lVert\rVert{}{
  \ifblank{#1}{\:\cdot\:}{#1}
}
\DeclarePairedDelimiter{\abs}{\lvert}{\rvert}

\DeclareUnicodeCharacter{00A0}{ }
\usepackage[frozencache, cachedir=_minted]{minted}
\newminted[she]{sh}{autogobble} 
\newmintinline[shl]{sh}{} 
\newminted[term]{console}{autogobble} 
\newmintinline[terml]{console}{} 
\newminted[ce]{c}{autogobble, fontsize=\small, tabsize=4} 
\newmintinline[cl]{c}{} 
\newminted[coqe]{coq}{autogobble,fontsize=\small} 
\newmintinline[coql]{coq}{} 

\newcommand{\eg}{e.g.}
\newcommand{\etal}{et al.}

\newcommand{\soft}[1]{\textsf{#1}}
\newcommand{\Coq}{\soft{Coq}}
  
  \newcommand{\Coquelicot}{\soft{Coquelicot}}

  \newcommand{\mathlib}{\soft{mathlib}}
  \newcommand{\MathCompAnalysis}{\soft{mathcomp-analysis}}

\newcommand{\Lean}{\soft{Lean}}

\newcommand{\eps}{\varepsilon}

\usepackage[normalem]{ulem}

\usepackage[color=orange]{todonotes}

\newcommand\USPN{Université Sorbonne Paris Nord}

\newcommand\LAGAaff{\USPN, CNRS,\\\LAGA,\\\USPNaddr}

\newcommand\LIPNaff{\USPN,\\\LIPN,\\\USPNaddr}
\newcommand\LIPNaffCNRS{CNRS, \USPN,\\\LIPN,\\\USPNaddr}
\newcommand\LMFaff{Université Paris-Saclay, Inria,\\%
  CNRS, ENS Paris-Saclay, LMF\\%
  91190 Gif-sur-Yvette, France}
\newcommand\Serenaaff{Inria,\\%
  48 rue Barrault, 75013 Paris.\\%
  CERMICS, École des ponts\\%
  77455 Marne-la-Vallée, France.}

\title{Maths with Coq in L1, a pedagogical experiment
  \thanks{%
    This work was partially supported by the Inria Challenge LiberAbaci {\url{https://liberabaci.gitlabpages.inria.fr/}}}}

\author{%
  Marie Kerjean
  \institute{\LIPNaffCNRS}
  \email{marie.kerjean@lipn.univ-paris13.fr}
  \and
  Micaela Mayero
  \institute{\LIPNaff}
  \institute{\LMFaff}
  \email{mayero@lipn.univ-paris13.fr}
  \and
  Pierre Rousselin
  \institute{\LAGAaff}
  \institute{\Serenaaff}
  \email{rousselin@univ-paris13.fr}
}

\def\authorrunning{Kerjean, Mayero, Rousselin}

\def\titlerunning{Maths with Coq in L1}

\begin{document}

\maketitle

\begin{abstract}
In France, the first year of study at university is usually abbreviated L1 (for
\textsl{première année de Licence}).
At ``Sorbonne Paris Nord'' University, we have been teaching an 18 hour
introductory course in formal proofs to L1 students for 3
years. 
These students are in a double major mathematics and computer science
curriculum. The course is mandatory and consists only of hands-on sessions
with the \Coq{} proof assistant.

We present some of the practical sessions worksheets, the methodology
we used to write them and some of the pitfalls we encountered.
Finally we discuss how this course evolved over the years and will see that
there is room for improvement in many different technical and pedagogical
aspects.
\end{abstract}

\section{Introduction}
\label{sec:intro}

In France, in most mathematics and/or computer science curricula, the first
semester after high school is often a key step for the students: the transition
to the ``rigorous
stage''\footnote{https://terrytao.wordpress.com/career-advice/theres-more-to-mathematics-than-rigour-and-proofs/}
of the mathematical activity, where the emphasis shifts from calculus to proofs.
This is done with different topics, of various abstract levels: \eg{} naive set
theory and relations, arithmetics, real analysis, ... Curiously, formal logic is
almost never considered, except for the occasional truth table, and the students
usually learn what is a proof by contradiction or a proof by induction without
knowing what a proof is.

In this context, we wanted to create at fall 2021 a new specific course for double major
mathematics and computer science students.
We wanted this new course to be
challenging, backed by research and at the interface of these two sciences.
We chose the \Coq{} proof assistant because we knew we could find
local technical expertise. The 18h constraint comes from the fact that the
course replaces an 18h methodology course and, in France, most computer science
departments are overloaded with work. To make the most of these 18h, we chose to
have only hands-on, 3h, sessions. Each year we had two groups of about 25
students each in computer labs.

This pedagogical experiment is not isolated, see \cite{assistants-preuves-smf}
for other computer assisted proof writing courses at the beginning of university
in France. Our course does not use, at this point, any layer on top of the proof
assistant. In this respect, it differs from other courses using GUI-based
software like Edukera or DEAduction.
Since we expect our students
to learn computer science and mathematics, we were not scared
to make them write proofs as sequences of tactics in a text editor. ``Software
Foundations'' by Pierce \etal{} (\cite{pierce2010software}) has been a key
influence. It certainly encouraged us to use plain \Coq{} so that the most
ambitious students could follow it after our course, without having to learn
basic proof writing a second time.

We decided quite early during the creation of the course that our ultimate goal
would be analysis of sequences of real numbers. On a pedagogical level, it is
very appealing since it involves many different quantifiers and we knew that
students usually struggle to prove correctly, \eg{} that the sum of two
converging sequences is converging (and even often have no idea what exactly
they are expected to prove).
In practice this was a bit too demanding for such a small course at the first
semester, but this will be discussed later.
With such a goal in mind, we settled each year for, more or less, the following
plan:
\begin{enumerate}
  \item Propositional logic
  \item Natural numbers and induction
  \item Predicate calculus
  \item Real numbers and sequences of real numbers
\end{enumerate}
The course, in its latest version, is available (in French) on a dedicated
\href{https://www.math.univ-paris13.fr/~rousselin/ipf.html}{webpage\footnote{\url{https://www.math.univ-paris13.fr/~rousselin/ipf.html}}}.
A read-only git repository with the $2023$ edition of this course, which is the
one we describe in this paper is available on
\href{https://github.com/Villetaneuse/ipf-2023}{GitHub}\footnote{\url{https://github.com/Villetaneuse/ipf-2023}}.
The course is (almost) completely centered on filling proofs for lemmas which
are already stated in the worksheets.

The rest of the paper is organized according to the same plan as the course. For
each of these items, we will sum up its content in our course and give some
feedback about its reception by the students. We will then explain our
assessment methods and will conclude with possible improvements in terms of
pedagogical and technical aspects.

Before, we briefly present other courses using a proof assistant at the early
years of university and how they differ from our own experiment. Patrick
Massot has created such a course at Paris-Saclay university. It also targets
double degree students in mathematics and computer science during their first
year, but during their second semester. His course uses \Lean{} with a set of
custom tactics called ``lean-verbose'' (\cite{massot2024teaching}) so that the
proof scripts written by the students are close to a (very detailed)
mathematical proof. His course is also directed towards elementary real analysis
and is backed by the community-developed
\href{https://github.com/leanprover-community/mathlib4}{\mathlib{}}\footnote{\url{https://github.com/leanprover-community/mathlib4}}
\Lean{} library for mathematics. At Université~Paris~Cité, Antoine Chambert-Loir
and Ricardo Brasca teach a purely \Lean{} and \mathlib{} based course (see
\href{https://plmlab.math.cnrs.fr/chambert/LeanTeaching}{its Gitlab
repository}\footnote{\url{https://plmlab.math.cnrs.fr/chambert/LeanTeaching}}
for its source code) which also covers linear algebra. Heather MacBeth's course
entitled ``\href{https://hrmacbeth.github.io/math2001/}{The Mechanics of
Proof}\footnote{\url{https://hrmacbeth.github.io/math2001/}}'' is another course
targeted at students learning how to write mathematical proofs, using again
\Lean{} with its \mathlib{}. It is oriented towards number theory, instead of
real analysis. Still with \Lean{} and the \mathlib{}, Jeremy Avigad's course
``\href{https://leanprover-community.github.io/mathematics_in_lean/}{Mathematics
in \Lean{}}\footnote{\url{https://leanprover-community.github.io/mathematics_in_lean/}}''
is of a much higher mathematical level, with topics such as linear algebra or
measure theory. Another \Lean{}-based experiment has been conducted by Frédéric
Tran-Minh (\cite{tranminh:hal-04705617}) comparing the difficulties of the
students when using ``term-mode'' \Lean{} \textsl{versus} the Edukera graphical
proof assistant.

In contrast to the \mathlib-based courses, our course uses an almost ``bare
bones'' \Coq, which has its pros and cons. Having less abstractions can be
appealing for a first course, especially for students learning both computer
science and mathematics. One may also be afraid that the various levels of
abstraction used in the \mathlib{} ``leak'', causing some tactics to have a
less predictable behaviour for the students, but we certainly need more insight
on this subject. On the other hand, it would be very difficult, for
instance, to cover linear algebra in this setting.

\section{Propositional logic: how to write an exercise sheet}
\label{sec:prop-log}
\subsection{Propositional (intuitionistic) logic}
We start with intuitionistic logic. The logical connectives
\coql!forall!, \coql!->! (for implication), \coql!/\! (for conjunction),
\coql!\/! (for disjunction) are described in a mechanical way by how we
can use them and how we can prove them.

We always start with a commented example. The students are expected to play
these examples in their interactive environment to see how each tactic modifies
the proof state. Our first ``Hello, World!'' example is \coql!imp_refl!:
\begin{coqe}
Theorem imp_refl : forall P : Prop, P -> P.
Proof.
  (* Let [P] be any proposition. *)
  intros P.
  (* To show an implication, one assumes that what is on the left of the arrow
     holds, and then prove what is on the right of the arrow. *)
  (* We assume (hypothesis ([H])) that [P] holds. *)
  intros H.
  (* We need to prove [P], but ([H]) is exactly a proof of [P]. *)
  exact H.
Qed. (* Quod erat demonstrandum. What was to be demonstrated. *)
\end{coqe}

Every year, showing this proof, during the first minutes of the first hands-on
session in September is a daunting ``throw the students
in the water'' moment.
We can't explain what is the type \coql!Prop!. They will need to get a feeling of what
it is by practicing. The students need to learn that
\coql!intros! may introduce a variable or an hypothesis. And there is even
Latin! It is worth noting that our first iteration was even worse: we thought it
would be a good idea to name the hypothesis \coql!HP!, but then many students
thought that the name of the hypothesis was meaningful and that calling it, say
\coql!HQ! would assume that some \emph{other} proposition \coql!Q! would
hold. We also used \coql!assumption! (which scans the context for a proof of the
goal) instead of \coql!exact H!, but this was far too magical at this stage.

That said, with practice, the students
actually manage to digest a lot more than what one would expect.
Following the example is (always) a straightforward exercise in order to ease
this digestion process.

And, from there, the students usually keep on working, with the occasional help
from the teacher.
In one or two sessions, almost all the students have learned the elimination
and introduction rules of the logical connectives (except \coql!exists! which is
seen later), how to read multiple arrows, \eg{} \coql!P -> (Q -> R)! as ``if \coql!P!
\emph{and} \coql!Q! hold then \coql!R! also holds'', ...

As in ``Software Foundations'', we have chosen our basic tactics to be as close
as possible to natural deduction rules; at the end of this exercise sheet
table~\ref{table:logical} is mostly complete.
 The \coql!False! proposition is
described by the principle of explosion, and the negation of a proposition,
\coql!~P!, is defined as \coql!P -> False!.
We keep away from the excluded middle at this point, because we feel that the
natural deduction rules are closer to usual mathematical practice (to prove an
implication, start with ``assume'').
We also restrict ourselves to backwards reasoning for the moment. This is
what \Coq{} natively encourages, and we have observed that exposing the students
to forwards reasoning too early (with \eg{} \coql!apply H in H'!) was a source
of confusion. Indeed, during this first exposure to the proof assistant, every
tactic should, ideally, fill one simple role. The implication to apply \emph{to
the goal} is chosen mostly by comparing the \emph{conclusion} of the implication with the
goal. When we introduce \coql!apply H in H'! too early, we add the rule that
we then need to compare the \emph{premise} of \texttt{H} with \texttt{H'}, which
may be one rule too much at this point. Another possibility, which we did not
try yet, would be to start with forward reasoning first, but most tactics modify
the goal anyway, so it may be unpractical.

\begin{figure}
\centering{}
\begin{tabular}{c|l|l}
  connective & introduction (prove) & elimination (use) \\ \hline
  $\rightarrow$ & \coql!intros! & \coql!apply! \\ \hline
  $\wedge$ & \coql!split! & \coql!destruct! \\ \hline
  $\vee$ & \coql!left! or \coql!right! & \coql!destruct! \\ \hline
  $\forall$ & \coql!intros! & \coql!specialize!,
  instantiation \\ \hline
  $\bot$ & & \coql!destruct! \\ \hline
  $\exists$ & \coql!exists! & \coql!destruct!\\ \hline
\end{tabular}
\caption{Logical tactics used in our course}\label{table:logical}
\end{figure}

\subsection{Sources of difficulties}
For the students, the main difficulty in this part is, when proving a
disjunction, the premature choice of \coql!left! or \coql!right!.
Some of them keep their eyes on their proofs (or worse, on the keyboard) and
do not watch closely enough the proof state to figure out that their proof
state is a dead end and they need to \emph{go back} in their proof.

The main difficulty as teachers, is to respect two guiding principles when
writing exercise sheets. First, the teachers have to honor a \emph{contract}: it
should \emph{always} be possible for a student to solve an exercise with the
tools (tactics, lemmas) which have been described beforehand. Second, teachers have
to do all they can to \emph{make the information flow manageable} by the
students. This last part can be very pedagogically challenging at times and
requires polishing again and again the exercise sheets. In some cases, it
exposed defects in \Coq{} itself and has lead to improvements. For instance, the
\Coq{} command \coql!Set Printing Parentheses!, which makes \Coq{} display, for
instance, \coql!A -> B -> C! as \coql:A -> (B -> C):, works as intended since
the recent version \texttt{8.19}, and a refinement to select the operators for
which we want parentheses displayed is under work.

A final important issue is the fact that we work with real-life \Coq{}, and
tactics shown in table~\ref{table:logical} \emph{go beyond the educational
framework} we installed, the biggest culprits being \coql!apply! (which is
a lot stronger than what we want it to be in this course) and
\coql!destruct! which can actually destruct almost anything in the context.
This is an issue when blocked students start trying things at random (and then,
when it magically works, share the solution with their classmates).
It then becomes the role of the teachers to enforce the rule:
\begin{center}
  \emph{Magic is not allowed in this course.}
\end{center}
and to discuss it with the students (what good is a solution you do not
understand?). It is tempting to write our own, more restricted, versions
of the basic tactics (with the same names to remain compatible with ``Software
Foundation''). We could then put back the real life versions when we are sure
that the basic logical rules are understood. This, however, is not a priority
at this point as it is not an insurmountable problem for the students.

\section{Natural numbers, computation and induction}
\subsection{Content of the worksheet}
Since we have a specific public and work at the interface of computer science
and mathematics, we chose to show how Peano's natural numbers
can be defined in \Coq{} as:
\begin{coqe}
Inductive nat : Set :=  O : nat | S : nat -> nat.
\end{coqe}
There is a little bit of hand-waving involved since we
do not go in the details of what an \coql!Inductive! type is. Instead, we
count on experiments and interactive feedback from \Coq{} itself to make
students understand the key concepts of ``no junk'' (all natural numbers
are obtained this way) and ``no confusion'' (if two natural numbers are
written differently they are different). It is also the first time the students
work with equality. We first present the \coql!discriminate! tactics which
proves \coql!False! from an equality such as \coql!S (S O) = S (S (S O))!.
Then we show that addition is defined as a recursive program in the following
way:
\begin{coqe}
Fixpoint add (n m : nat) : nat :=
  match n with
  | 0 => m (* If n = 0, then n + m = m. *)
  | S p => S (add p m) (* (S p) + m = S (p + m). *)
  end.
\end{coqe}
and students are then encouraged to use pen and paper to compute $2 + 2$ using
these two rules.
This approach, which is closer to computer science than mathematics, has some
interests (especially in our context) but it has some drawbacks we will discuss
shortly. We then introduce the \coql!reflexivity! tactics which, in our course,
is used to prove an equality when both sides are syntactically
identical\footnote{This is an oversimplification and \coql!reflexivity! goes
  beyond our framework for two reasons: first, it checks (computes) that both
  sides are convertible, second, it is tweaked with typeclass instances, so that
it can prove, for instance, that $x \le x$.}. We show how the \coql!simpl! tactic asks \Coq{}
to use the addition program as much as it can in a term.
Students then use \coql!destruct! to perform proofs by case on the nullity of an
integer and \coql!rewrite! to \emph{use equalities} to replace some terms in the
goal with other terms.

At this point, the students are shown that proofs by cases are not sufficient to prove,
for instance, that $\forall (n : \mathbb{N}), n + 0 = 0$, which allows us to
introduces the \coql!induction! tactic. This tactic, and how it forces the
students to prove a base case and an induction step with an induction hypothesis
is arguably enough to legitimate this course by itself.
With it, students can sharpen a skill which is essential both in computer
science and mathematics. From there, the multiplication is defined and students
prove basic properties of addition and multiplication such as commutativity,
associativity, etc. As a good side effect, it's also a very concrete exposition
to these algebraic notions which are mostly new for the students.

\subsection{Activity}

At this point, we would like to insist on a key aspect of this course: \emph{the
students are constantly active.} Many of them even skip the 15 minutes break and
work without interruption for 3 hours. We believe that there are three reasons
to that.

First, the interactive aspect of the proof assistant, giving \emph{instant
feedback}
and allowing to experiment. The quick \coql!Qed.! reward certainly also plays a
role. Some students compared the course to a video game. Second, the
joy of theorem proving itself (with or without a proof assistant) should not
be neglected, \emph{it is made easier by the fact that everything here is
explicit.} There is no hidden forgotten sine identity trick needed to solve
an exercise. Finally, we put a lot of efforts to \emph{make the difficulty
as progressive as it could be.} Of course, if students were facing an impossible
task from the beginning, we would lose the benefits of the first two points.

\subsection{Computational aspects and rewriting}

There are still some points of friction and possible improvements (or different
choices). First, the computational aspects are not well understood.
Some students try to use \coql!simpl! or \coql!discriminate! at random.
The \coql!simpl! tactic often gives more than what
was expected. For instance, using \coql!simpl! on the goal
\begin{coqe}
(1 + n) ^ 2 = n ^ 2 + 2 * n + 1
\end{coqe}
gives
\begin{coqe}
S (n * 1 + n * S (n * 1)) = n * (n * 1) + (n + (n + 0)) + 1
\end{coqe}
which then makes the proof almost impossible (and remember that students rarely
go back to make different choices). This year, we have
restricted the usage of \coql!simpl! to proofs of basic rewriting rules derived
from the definition and encouraged the students to use the \coql!rewrite! tactic
instead.
About the computational aspects themselves, we think we
are sitting on a fence between either dropping them completely (and then
presenting \coql!nat! in an axiomatic way) or putting more time and effort to
it. With our specific public, we would like to try the second approach in the future.
In a course for mathematics students only, we would recommend the first one.

Finally, \coql!rewrite! has its own issues. It can be hard to select precisely
the part of the goal one wants to use, say, commutativity on.
We could introduce
\coql!replace! at this point, but this would introduce yet another tactic, and
it makes proofs more verbose. Another option, when rewriting becomes very tedious,
would be to introduce more automation with tactics such as \coql!ring! and
\coql!lia!. We opted out of it at this point. These powerful automation
tactics should be introduced only when it is clear that they only offer time
saving. Otherwise we risk to encourage magical thinking.

\section{Predicate calculus and classical logic}
\subsection{Content of the worksheets}
The predicate calculus worksheet aims at working with \coql!forall! and
\coql!exists! predicates. The \coql!specialize! tactic plays the role of
an elimination rule for the \coql!forall! predicate. We work both with
abstract formulas, for instance:
\begin{coqe}
Theorem forall_or_forall :
  (forall x, (P x)) \/ (forall x, (Q x)) -> forall x, (P x \/ Q x).
\end{coqe}
and concrete predicates, for instance \coql!is_null! and \coql!is_non_null! on
\coql!nat! to provide examples and counter-examples.

We believe that spending time on the \coql!exists! connective is very important.
At this point, there is still no excluded middle, so the only way to prove
an existential formula is to exhibit a witness with the \coql!exists! tactic.
The elimination of such a formula is also interesting since it introduces
in the context a new element with only one assumption about it. The students
spend a lot of time working on predicate calculus tautologies or false formulas.
We make it apparent that \coql!exists! behaves like a generalized \coql!\/!
and \coql!forall! behaves like a generalized \coql!/\!.

Next is a small worksheet about classical logic. We add the excluded middle and
derive new logical identities which were not provable without it. In particular,
we work on negations of existential and universal formulas.

\subsection{A very apparent speed decrease}

The students usually find these two worksheets a lot more difficult than the
previous ones. Everything takes a lot of time. We have (non measurably)
identified two causes. First, some students have difficulties to use their
mathematical intuition in front of the proof assistant. The relation between
formal proving and their usual mathematical activity is not clear to them. We
hope to work on this aspect in the next course sessions. Second, they struggle
with more abstract, higher-order logic. This might be a mathematical issue,
where the proof assistant acts as an amplifier of the students difficulties.
There is certainly no easy solution to this pedagogical problem, which is
probably as old as mathematics. We intend to try adding intermediate steps, for
instance predicates on a finite type.

We were also surprised by the difficulties of the students in classical logic.
After all, it's ``their'' logic in everyday mathematics. It turns out that they
do not relate strongly truth tables and proofs. Even when this is is not an
issue, the excluded middle is very special since, in contrast with other logical
rules, it can happen any time with any proposition. On a more technical side,
at this point \Coq{} with its standard library does not offer much to work
with classical logic. For instance it lacks an equivalent of the \mathlib{}'s
\href{https://leanprover-community.github.io/mathlib4_docs/Mathlib/Tactic/PushNeg.html}{\texttt{push\_neg}\footnote{\url{https://leanprover-community.github.io/mathlib4_docs/Mathlib/Tactic/PushNeg.html}}}
tactic which transforms automatically a negated first-order formula with an
equivalent one where the negation is at the end.
This should be improved if one wants to use it more for
mathematical courses.

\section{Real numbers}

\subsection{Content of the last worksheets}

For real numbers, as in the rest of the course, we use \Coq{}'s standard library.
Historically, \Coq'{}s real numbers have been introduced axiomatically
by Micaela Mayero (\cite{May01}) as an ordered field satisfying the
least-upper-bound property. They have been constructed more recently by Vincent
Semeria (\cite{semeria2020nombres}) using Dedekind's cuts\footnote{Actually,
  there are now two flavors of real numbers, one is constructive, the other not;
we work with the non-constructive version.}. We do not want to expose the
internal construction to the students, so we present the real numbers in an
axiomatic way. This has the good side effects to introduce them to abstract
algebra.

The operations and constants are introduced progressively, together with the
axioms they satisfy and consequences of these axioms. For instance, we ask
them to prove
\begin{coqe}
Theorem unique_opp : forall x y z : R, x + y = 0 /\ x + z = 0 -> y = z.
\end{coqe}
using the fact that $(\mathbb{R}, +)$ is a commutative monoid.
From the axioms, we play the ``real numbers game'' until we prove that
$0 < 1$. In passing, the students also work with strict and non-strict orders
and compatibility properties. Many proofs involve transitivity and
antisymmetry, on top of the other, algebraic, properties of the operations.

The next exercise sheet deals with the absolute value. It was added in 2022: we
saw during the first edition of the course that the students were not familiar
enough with this notion to handle subsequent real analysis work.

Finally, the worksheet about real sequences introduces the convergence of real
sequences. The example shown is the uniqueness of the limit. Our initial goal
for this whole course, was that the students prove the following lemma (given,
here, with its solution):

\begin{coqe}
Theorem CV_plus (An Bn : nat -> R) (l1 l2 : R) :
  Un_cv An l1 -> Un_cv Bn l2 -> Un_cv (fun n => An n + Bn n) (l1 + l2).
Proof.
  unfold Un_cv.
  intros HA HB eps Heps.
  destruct (HA (eps / 2)) as [n1 Hn1]. lra.
  destruct (HB (eps / 2)) as [n2 Hn2]. lra.
  remember (max n1 n2) as n3 eqn:def_n3.
  exists n3.
  intros n Hn.
  replace eps with (eps/2 + eps/2) by lra.
  apply (Rle_lt_trans _ ((R_dist (An n) l1) + (R_dist (Bn n) l2))). {
    apply R_dist_plus.
  }
  apply Rplus_lt_compat.
  - apply Hn1. lia.
  - apply Hn2. lia.
Qed.
\end{coqe}
Even without prior \Coq{} experience, one can probably see how it closely
matches the usual mathematical proof. There are, however, subtle points which
need be discussed in order to grasp the difficulties of the students.
After the introductions, the hypothesis \coql!HA! has type
\begin{coqe}
HA : forall eps, eps > 0 -> exists N, forall n, n >= N -> (Rdist (An n) l1) < eps
\end{coqe}
What we perform in the first \coql!destruct! tactic is first, to specialize it
with \coql!eps / 2!, then extract a witness \coql!n1! satisfying the property
\coql!Hn1!, \emph{and add the new subgoal} that \coql!eps / 2 > 0!.
This new subgoal would certainly never be explicit in a mathematics course.
To handle such cases, we have allowed the usage of the powerful solvers
\coql!lra! (for ``Linear Real Arithmetic'') and \coql!lia! (for ``Linear
Integer Arithmetic). In our case, \coql!lra! can handle the proof that
\coql!eps / 2 > 0! without complaining.

The next step is to consider the maximum \coql!n3! of the two indices \coql!n1! and
\coql!n2!, this is done with the tactic \coql!remember!.
The most tedious part to write in Coq is certainly the use of the triangle
inequality. In a mathematical course, this would probably have been written
\begin{align*}
    \abs{A_n + B_n - (l_1 - l_2)} &= \abs{A_n - l_1 + l_2 - B_n }\\
                                  &\le \abs{A_n - l_1} + \abs{B_n - l_2}
                                  &&\text{(by the triangle inequality)}\\
                                  &< \frac{\eps}{2} + \frac{\eps}{2} = \eps,
\end{align*}
where one does not need explicitly stating transitivity results.
We would be happier if our formal proof resembled this. It is not possible in
our course yet,
but \Lean{} has the \coql!calc!
tactic\footnote{https://leanprover-community.github.io/extras/calc.html}
which does just that and
\texttt{coq-waterproof}\footnote{https://github.com/impermeable/coq-waterproof}
also allows this style, so we can reasonably hope this will be available in the
near future.
Notice the usage of the \coql!lia! tactic in the last steps, to show
that, indeed, \coql!n3! is greater than \coql!n1! and \coql!n2!.

Now, honesty compels us to say that very few students actually managed to prove
this theorem (only 5 students in 2022, none during the two other sessions).
In 2021, it was our first time writing and teaching this course, so it was a bit
less polished. In 2023, we made a big pedagogical mistake about forward reasoning
(more about this later) which made the student lose a lot of time.
Anyway, 18h is probably not enough to cover that much mathematical and proof
assistant related content during the first semester of university.

\subsection{Forward reasoning, \texttt{rewrite} and \texttt{apply}}

As mentioned earlier, \Coq{} has a tendency towards backwards reasoning.
One possible reason is that there is always one active goal and maybe many
hypotheses, so this is a bit simpler to write.
Still, it is possible to \coql!apply! an implication, say a proof
\coql!HI! of \coql!A -> B! \emph{in} a proof, say \coql!H!, of some
proposition \coql!A! to transform \coql!H! into a proof of \coql!B!.
It is arguably the more natural way to write a mathematical proof: start with
the assumptions, and derive consequences until you reach the goal.

In $2021$, we introduced this style in the first exercise sheet, alongside with
backwards reasoning and we have observed that this created some confusion for
the students, who already had a lot to digest.
In $2022$, we have introduced this style later, just before working on real
numbers. In $2023$, we thought that (almost) removing forward reasoning would
simplify our exposition, hence the life of our students. In the same movement,
we omitted such lemmas as
\begin{coqe}
Lemma Rplus_eq_compat_l : forall r r1 r2 : R, r1 = r2 -> r + r1 = r + r2.
\end{coqe}
which have no mathematical content and serve primarily in forward reasoning. But
then, we noticed that they were quite slower in the part about real numbers
(without at first, identifying why).

It turns out dropping forward reasoning was a big mistake. \emph{We should not prevent
the students to write their proofs in the way they are used to.} On the
contrary, the proof assistant should get closer to what is deemed more natural
to the user. Still, the risk to create confusion between forward and backward
reasoning exists and we do not know properly yet when and how to discuss this
distinction in our course.

Another source of difficulty is the confusion between \coql!rewrite! and
\coql!apply! when the conclusion of the implication is an equality.
For instance, consider the following cancelation lemma:
\begin{coqe}
Lemma Rplus_eq_reg_l: forall r r1 r2 : R, r + r1 = r + r2 -> r1 = r2.
\end{coqe}
and assume we want to show:
\begin{coqe}
Theorem double_fixpoint_0 : forall (x : R), x + x = x -> x = 0.
Proof.
  intros x H.
\end{coqe}
At this point, using \coql!apply (Rplus_eq_reg_l x)! unifies \coql!r1! with
\coql!x! and \coql!r2! with \coql!0! and changes the goal into the manageable
\coql!x + x = x + 0!. This kind of reasoning requires that the user have
acquired some reflexes concerning unification (and again, it is probably more
natural for our students to transform the hypothesis \coql!H! into
\coql!x + x = x + 0! and then apply \coql!Rplus_eq_reg_l! \emph{in} \coql!H!
without having to instantiate any variable explicitly).
However, if the student tries to \coql!rewrite Rplus_eq_reg_l!, \Coq{} will
rightfully complain that it has no idea which term to rewrite (\coql!r1!) and
into what (\coql!r2!). Even if, after struggling, the user finally manages
to instruct \coql!rewrite (Rplus_eq_reg_l x x 0)!, then \Coq{} will change
the goal into \coql!0 = 0! and ask for a proof that \coql!x + x = x + 0!, which
is slightly worse than the \coql!apply! solution.

In practice, we noticed that this distinction between \coql!apply! and
\coql!rewrite! is not well understood and we should probably take more time
to work on it with the students.

\subsection{Solvers}
As we mentioned earlier, we introduce the tactics \coql!lia! and \coql!lra! very
late in the course. We feel that the students need to be able to prove
most easy equalities and inequalities, before giving them these powerful
solvers. However, it is clear that they also serve pedagogical purposes: with
them it is possible to treat exercises that would have been be too tedious
otherwise.
This year, the last exercise of a homework allowed the students to use about
anything at their disposal. Here is the statement, with a part of a student's
solution.
\begin{coqe}
Lemma polynome2_positive (a b c : R) : a > 0 -> (b² - 4 * a * c < 0) ->
  forall x, a * x² + b * x + c > 0.
Proof.
  unfold Rdiv,Rsqr.
  intros H H1 x.
  replace (a*(x * x) + b * x + c) with
    (a*((x+(b*/(2*a)))² +(- (b²-4*a*c))*/(4*a²))).
  - apply (Ropp_gt_lt_0_contravar ((b²-4*a*c))) in H1.
    (* 9 lines to prove that a*((x+(b*/(2*a)))² +(- (b²-4*a*c))*/(4*a²)) > 0 *)
  - unfold Rsqr. field. lra.
Qed.
\end{coqe}
As one can see, the heart of the proof is the transformation of the polynom
into its canonical form. This was not guided, so the student took this
initiative herself, and then, \emph{after writing the equation}, used the
\coql!field! tactic \emph{to verify it}.

\section{Assessments}
The students have to upload two homework assignments (as \Coq{} files) and take a final
1h30 exam in computer lab.

The homework assignment have easy exercises as well as challenging ones. Except the
first year, there has always been a handful of perfect homework assignments.
We adopt a very simple grading scheme with \coql!Qed! or nothing except in very
specific edge cases. These are quickly corrected with the help of \Coq{} (and a
quick human read to watch for edge cases). We should certainly add more
homework assignments in the future.

We did not want to make this course too centered around assessments (in part
because it is still somewhat experimental), so its
weight in the final semester grade is quite small (1 out of 30).
Moreover, it is good to tell the students once in a while that they are not here
to learn how to pass exams but to study mathematics and computer science.
Still, students being students, they take their grades very seriously.
We wanted the grades to be ``good'', in a French sense, so with a mean of
around 14/20, in order to not handicap our double major students (which would
certainly get good grades in the methodology course it replaces).

In the end, the heterogeneity of the students is incredible. A few of
them still struggle with basic logic, while others are able to prove more than
twenty lemmas, some of them not that trivial, during the exam. 
A strong majority of students solve all basic logic exercises and most easy
inductions. We have opted for
a diminishing returns scale : the first \coql!Qed! weighs twice more than the
tenth which weighs twice more than the twentieth.

\section{Conclusion and prospects}

We have shown the content of our course and emphasized some of the difficulties
we, or the students, have encountered, be them of pedagogical or technical
origins. Our course contrasts with other such experiments in its usage of
``real life (plain) \Coq'' instead of a specific layer on top of it, for better
and for worse. There is room for improvement in many different directions.
We already mentioned that we could work on restricting the use cases of our
tactics to their pedagogic ideal, but we are not experts on this subject.
The students could also benefit from a rewriting of the \coql!Reals! library,
so that they face a more uniform naming scheme. Other, more modern, libraries
deal with analysis: \Coquelicot{} (see \cite{boldo2015coquelicot}) and \MathCompAnalysis{}
(see \cite{affeldt2018formalization}) offer a lot more, but are not really
suited for teaching at this level since they use more advanced and abstract
notions such as abstract algebraic hierarchies and filters.

However, the biggest improvements, from our perspective, would be about user
interfaces in a broad sense : in terms of both the input/output of \Coq{} itself
and graphical interface.
From \Coq{}, we mentioned the need to ease writing of equalities and
inequalities in general. Error messages and the absence of hints are also a
concern. As for the graphical interface, we have been using \texttt{coqIDE} in
2021 and 2022 and have switched to \texttt{jsCoq} in 2023.
Using \texttt{jsCoq} is a lot better: the exercise sheets are opened in a Web
browser and the comments are displayed as HTML. The
students actually read the course while, with a text editor, they usually
skipped the commentaries, displayed as a greyish heap of ASCII characters.
We could have a lot more, for instance a list of authorized lemmas for each
exercise, so that the student can browse a manageable list of results at any time, or
easy to write mathematical formulas, function plots, ...
In short, since \Coq{} is a member of the teaching team, we would like to be
able to tweak it, in order to make it a better teacher.

Finally, we need to assess the usefulness of this course on the mathematics
side. The course certainly does not harm (the results of our students in
mathematics courses are quite good) but it would be useful to have more specific
feedback in that respect.
Starting next year, we will add pen and paper exercise
sessions to work on mathematical proof writing in relation to the work with the
proof assistant. This will certainly help in this regard, but this assessment
task is certainly difficult. First, it is hard to measure the gap between the
activities of typing formal proofs with \Coq{} and writing mathematical
pen-and-paper proofs to convince human beings. Then, we would need to understand
better which part of this gap is related to \Coq{} and/or our formalization
choices for this course. The recent paper \cite{bartzia:hal-04087080} presents a
first analysis of the differences between proof assistants for education.
There is certainly a lot more to study in this direction.

\section*{Acknowledgments}
The authors would like to warmly thank our double major students from 2021 until
now, for being enthusiastic learners and giving relevant and useful feedback.


\bibliographystyle{eptcs}
\bibliography{biblio}

\end{document}